\documentclass[10pt, letter, conference]{IEEEtran}
\usepackage{cite}
\usepackage{amsmath,amssymb,amsfonts}
\usepackage{algorithmic}
\usepackage{graphicx}
\usepackage{textcomp}
\usepackage{xcolor}
\usepackage{tikz}
\usepackage{subcaption}
\usepackage{enumitem}
\usepackage{array}
\usepackage{todonotes}
\usepackage{url}
\usepackage{breakurl}
\usepackage{silence}
\ErrorFilter{caption}{table inside subfigure}

\usepackage{listings}
\def\BibTeX{{\rm B\kern-.05em{\sc i\kern-.025em b}\kern-.08em
    T\kern-.1667em\lower.7ex\hbox{E}\kern-.125emX}}
\newcommand{\sqnum}[1]{%
  \tikz[baseline=(n.base)]{%
    \node[
      fill=black,
      text=white,
      minimum width=1.05em,
      minimum height=1.05em,
      inner sep=0pt,
      font=\footnotesize\bfseries
    ] (n) {#1};%
  }\,%
}
\newcommand{\sqall}{%
  \tikz[baseline=(n.base)]{%
    \node[
      fill=black,
      text=white,
      minimum width=1.05em,
      minimum height=1.05em,
      inner sep=0pt,
      font=\footnotesize\bfseries
    ] (n) {All};%
  }\,%
}
\newcolumntype{C}[1]{>{\centering\arraybackslash}m{#1}}
\newcolumntype{L}[1]{>{\raggedright\arraybackslash}m{#1}}

\IEEEoverridecommandlockouts\IEEEpubid{\makebox[\columnwidth]{979-8-3195-0662-7/26/\$31.00 $\copyright$2026 IEEE \hfill}\hspace{\columnsep}\makebox[\columnwidth]{ }}

\usepackage[absolute,overlay]{textpos}

\begin{document}

\title{Extending Low Latency Service Across the Internet
}

\author{
\IEEEauthorblockN{
Harkirat Singh\textsuperscript{1,*},
Fatih Berkay Sarpkaya\textsuperscript{2,*},
Hakan Gulec\textsuperscript{2},
Fraida Fund\textsuperscript{2},
Shivendra Panwar\textsuperscript{2}
}
\IEEEauthorblockA{
\textsuperscript{1}Stony Brook University, NY, USA;
\textsuperscript{2}NYU Tandon School of Engineering, Brooklyn, NY, USA\\
\{hs5076, fbs6417, hg2918, ffund, panwar\}@nyu.edu
}
}

\maketitle

\begin{abstract}
Protocols such as L4S for low latency network services have attracted growing interest from major industry stakeholders such as Comcast, Apple, T-Mobile, and NVIDIA. However, L4S requires isolation between L4S flows and classic flows in order to maintain its low latency benefits and safe coexistence. The L4S architecture uses a DualPI2 AQM mechanism to provide this isolation when the bottleneck occurs at the home access link. In practice, however, bottlenecks may also occur at other locations in the network, such as peering points, ingress to wide area networks, or occasionally congested links in the core, where deploying DualPI2 AQM is not feasible. Without a solution to this problem, L4S may struggle to gain deployment, since its benefits may remain limited without end-to-end support. To address this challenge, we propose a deployment strategy that reduces the need to upgrade expensive core routers by using existing networking mechanisms such as BGP communities, SRv6, and priority queues with rate caps. We demonstrate the feasibility of this approach through large scale experiments on the FABRIC testbed. Our results demonstrate a practical deployment strategy for extending low latency service across the Internet, even across independently administered networks.
\end{abstract}


\begingroup
\renewcommand{\thefootnote}{}
\footnotetext{%
\hspace*{-1.15em}\textsuperscript{*}Equal contribution.%
}
\endgroup

\begin{textblock*}{0.83\paperwidth}(18mm,258mm)
\scriptsize
\noindent \textcopyright{} 2026 IEEE. Personal use of this material is permitted. Permission from IEEE must be obtained for all other uses, in any current or future media, including reprinting/republishing this material for advertising or promotional purposes, creating new collective works, for resale or redistribution to servers or lists, or reuse of any copyrighted component of this work in other works.
\end{textblock*}

\section{Introduction}

Low latency networking is increasingly important for interactive and real time applications, including AR/VR, cloud gaming, video calling, and remote control systems. The main barrier to deployment of low latency transport at Internet scale is not the absence of candidate protocols, but limited deployability across heterogeneous end-to-end paths. Over the past decade, the community has proposed and evaluated multiple approaches. In particular, L4S has emerged as an architectural approach for low latency service that combines scalable congestion control, accurate ECN signaling, and dual queue bottleneck treatment~\cite{l4sarch-rfc9330,l4s-ecn-rfc9331,dualpi-rfc9332}. While this approach shows strong results in controlled evaluations and early deployments~\cite{livingood-low-latency-deployment-16}, its behavior is less predictable on heterogeneous end-to-end paths and in mixed traffic conditions, as illustrated by coexistence results from recent L4S studies~\cite{to-switch-or-not-to-switch-prague-sarpkaya,pam-sarpkaya} and real deployment measurements showing some benefit in a single domain~\cite{pam-2026-feamster-L4S}.

At the core is a coexistence problem at shared bottlenecks. Scalable congestion controls such as TCP Prague rely on shallow queues and fine-grained congestion signals applied at a low threshold, while classic loss-based transports such as CUBIC and Reno depend on deeper queues and higher marking thresholds to sustain utilization. When these traffic classes share a bottleneck queue, their control objectives conflict. Prior work shows the sensitivity of fairness outcomes to shared bottleneck conditions \cite{to-switch-or-not-to-switch-prague-sarpkaya,pam-sarpkaya}.

Dual queue AQM is the L4S response to this problem. It separates low latency and classic traffic while coupling queue behavior to preserve utilization and fairness. This approach performs well when deployed at the active bottleneck~\cite{to-switch-or-not-to-switch-prague-sarpkaya,pam-sarpkaya}. In practice, however, end-to-end paths include many potential congestion points, some of which are impractical to upgrade with dual queue support. Deployment becomes even more challenging across multiple administrative domains. Even when one domain provides isolation, without interdomain coordination, low latency guarantees can break at domain boundaries. This motivates our focus in this paper: a deployable strategy that preserves low latency service across multiple bottleneck locations and domains using mechanisms that operators can incrementally deploy.

In this work, we design and evaluate a strategy combining three mechanisms: priority isolation with rate caps at shared border bottlenecks, SRv6 path steering inside upgraded domains, and BGP community signaling to prefer interdomain paths that support low latency service. We evaluate this design on the FABRIC testbed with controlled bottleneck placement and realistic mixed workloads. Our results show that the strategy largely restores L4S application performance when congestion appears at sender access, peering, and core locations, while maintaining acceptable impact on classic traffic.

The contributions of this work, therefore, are as follows:

\begin{itemize}
\item We demonstrate that the scale of L4S deployment may be limited due to poor performance when the bottleneck is not at the receiver access link.
\item We propose a practical deployment strategy for L4S that minimizes the extent of required updates to core Internet routers, while preserving end-to-end low latency service across different bottleneck locations.
\item We evaluate the proposed strategy in a series of systematic experiments on the FABRIC testbed.
\item We make all experiment materials available so others can validate and build on our work.\footnote{%
Artifacts are available at: \url{https://github.com/fatihsarpkaya/L4S-ICNP2026}
}
\end{itemize}

The rest of this paper is organized as follows. Section~\ref{sec:background} provides background on L4S and reviews related work. Section~\ref{sec:solution-description} describes our proposed strategy for practical deployment of L4S across the Internet. Section~\ref{sec:methodology} details the experiment methodology. Section~\ref{sec:results} presents the experiment results. Section~\ref{sec:limitations} notes some limitations, and Section~\ref{sec:conclusion} concludes.

\section{Background and Related Work}
\label{sec:background}

Prior work shows that L4S can provide low latency when its main components are deployed together, especially when the bottleneck is at the access network. However, these benefits remain less certain under partial deployment, as L4S flows may encounter non-L4S bottlenecks or classic competing traffic elsewhere on the end-to-end path. This section reviews the L4S architecture, deployment and measurement studies, coexistence issues, and congestion beyond the access edge.

\subsection{L4S Architecture}

The L4S architecture, defined in RFC~9330, RFC~9331, and RFC~9332~\cite{l4sarch-rfc9330,l4s-ecn-rfc9331,dualpi-rfc9332}, is designed to provide low queuing delay, low loss, and scalable throughput. It relies on three main components: scalable congestion control at the sender, accurate ECN feedback from the receiver, and a queue that supports L4S at the bottleneck. TCP Prague~\cite{briscoe-iccrg-prague-congestion-control-04} is the reference scalable congestion control, while AccECN~\cite{rfc9768-accecn} provides the accurate congestion feedback needed by scalable senders. At the bottleneck, DualPI2 separates L4S and classic traffic into two queues: L4S traffic, identified by the ECT(1) codepoint, enters the low latency queue, while classic traffic enters the classic queue, and the queues are coupled to enable low latency for L4S while sharing capacity with classic traffic. When all components are present, L4S delivers very low queuing delay without sacrificing throughput. In practice, however, an L4S flow may traverse bottlenecks that do not support L4S, or may compete with classic flows that do not use scalable congestion control, making partial deployment an important challenge.

\subsection{L4S deployment}

L4S is moving toward production deployment. Comcast reports deploying dual queue mechanisms at scale across its residential broadband network, with measured reductions in loaded latency for both L4S and Non-Queue-Building (NQB) traffic~\cite{livingood-low-latency-deployment-16}. T-Mobile reports deploying L4S in its wireless network~\cite{tmobile_l4s_5g_advanced_2025}; Apple supports L4S in FaceTime~\cite{apple-l4s-debug}; and NVIDIA GeForce NOW supports L4S on compatible networks and devices~\cite{nvidia}. Together, these deployments demonstrate the growing adoption of L4S for applications sensitive to latency, such as video calls, cloud gaming, and interactive media.

Recent work measures Apple services over Comcast’s residential broadband network~\cite{pam-2026-feamster-L4S}, and compares L4S-enabled FaceTime, iCloud, and Apple CDN traffic with non-L4S baselines. It finds that L4S reduces tail latency for interactive traffic and bulk downloads, particularly under congestion, but provides limited benefit for short, bursty iCloud sync traffic.

Comcast’s deployment at scale places dual queue support near subscriber bottlenecks in the access network~\cite{livingood-low-latency-deployment-16}, a practical starting point because many delays visible to users arise there. However, this approach does not address congestion elsewhere along end-to-end paths. Controlled experiments show that TCP Prague achieves low queuing delay and coexists fairly with CUBIC at a DualPI2 bottleneck~\cite{should-we-care-L4S}, but do not establish its behavior when congestion occurs at bottlenecks without DualPI2. The production study likewise neither isolates bottleneck locations nor controls cross traffic~\cite{pam-2026-feamster-L4S}.

This suggests an important deployment gap. An end-to-end Internet path may include sender access links, peering links, transit domains, and core links that do not support DualPI2. Upgrading every such bottleneck is unlikely to be practical across multiple administrative domains. A deployable L4S strategy should therefore preserve low latency service beyond the receiver access link while minimizing the number of routers requiring L4S specific upgrades. This paper addresses that gap by combining isolation at shared bottlenecks, path steering inside upgraded domains, and interdomain signaling to identify paths that support low latency service.

\subsection{L4S Coexistence in Partial Deployment}

Several studies have evaluated how L4S behaves when one or more of its components are missing. Early experiments~\cite{l4s-tests-heist,henderson2019l4s-l4s-testing} showed that L4S traffic can dominate classic traffic when both share a single ECN-enabled queue. TCP Prague includes an ECN fallback heuristic~\cite{prague-fallback} to reduce this problem, though reliably detecting such bottlenecks remains difficult.
More recently,~\cite{to-switch-or-not-to-switch-prague-sarpkaya,pam-sarpkaya} evaluate congestion controls compatible with L4S against CUBIC and BBR variants across different bottleneck queue types. These studies find that safe coexistence is not guaranteed: performance depends strongly on the bottleneck queue and the competing congestion control, and the full low latency benefit of L4S is realized only when all key components are in place.

\subsection{Congestion points}

There is widespread agreement that congestion can occur in customer access networks, but queueing may also arise beyond the access edge. Prior work identifies several possible congestion points, including intra-provider links, interdomain peering and interconnection links, and transit or WAN-ingress paths used by large cloud and content providers.

Early wide-area measurements~\cite{10.1145/781027.781075} show that bottlenecks can appear both within ISP networks and between neighboring ISPs. Similarly,~\cite{19826} found that congestion location differs across access technologies: DSL subscribers were more often last-mile limited, while cable subscribers typically experienced congestion in the ``middle mile'' or beyond.

Peering, transit, and WAN ingress links can also become operationally significant congestion points. One measurement study found little evidence of persistent congestion on most interdomain links it examined, but observed prolonged congestion on a small number carrying high-volume traffic from specific providers~\cite{2014-clark-maii}.  Other studies likewise find recurring congestion on specific interdomain links~\cite{10.1145/2663716.2663741,10.1145/3230543.3230549}. An analysis of video conferencing traffic~\cite{10.1145/3696404} found patterns consistent with congestion at transit ISPs, and~\cite{10.1145/3544216.3544234} noted that some workloads can overwhelm individual WAN ingress links.

\begin{figure}[t!]
    \centering
    \includegraphics[width=0.8\columnwidth]{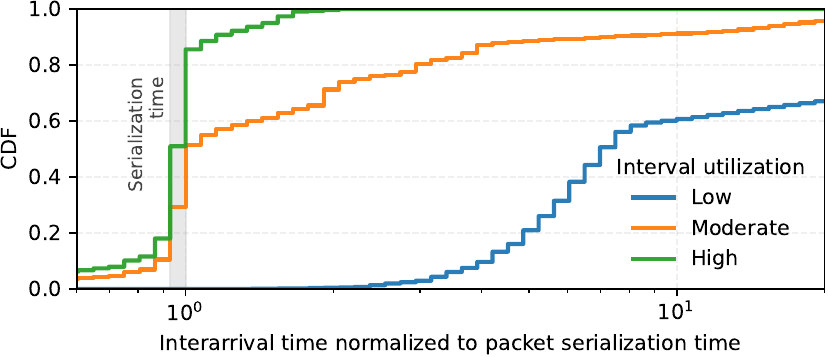}
    \caption{CDF of packet inter-arrival time normalized to the packet serialization time on a 100~Gbps backbone link. Although the link is not fully utilized, a substantial mass of packets arrives at exactly the serialization time, suggesting queuing at or upstream of the monitor.}
    \label{fig:interarrivals}
     \vspace{-1em}

\end{figure}

For additional evidence that queueing occurs beyond the access edge, we analyze CAIDA's Anonymized Two-Way Traffic Packet Header Traces~\cite{caida100g} from a 100~Gbps backbone link, which has layer 1--4 packet headers from passive traces captured on high-speed backbone monitors; we use data from November 2025 on the Los Angeles--Dallas link.
Figure~\ref{fig:interarrivals} shows the CDF of packet inter-arrival times normalized to their serialization time on the 100~Gbps link. Under moderate or high utilization, we note a substantial mass of packets arriving at exactly the serialization time, implying that many packets experience queuing at or upstream of the monitor. (There is also a small mass of packets arriving slightly sooner than the serialization time; we determined that it is composed of small packets whose normalized inter-arrival time is disproportionately affected by a few nanoseconds of timestamp noise.) This supports the view that queueing can occur beyond the customer access network, including on backbone links.

\subsection{Experiment: Upstream Congestion Beyond the Access Link}

\begin{figure}[h]
    \centering
    \includegraphics[width=\columnwidth]{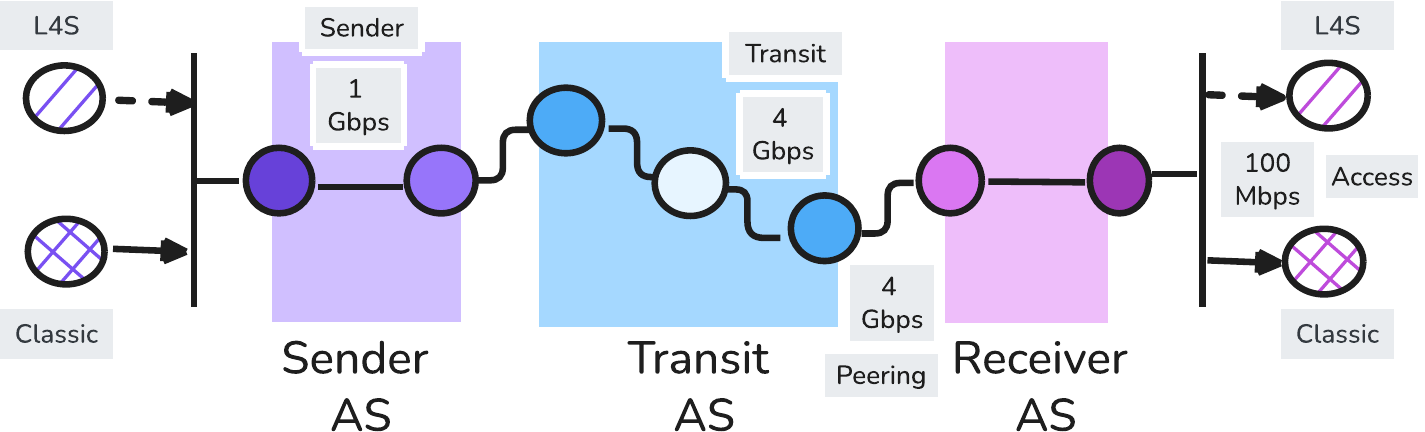}
    \caption{Topology for the motivating experiment. A bottleneck may occur at the receiver access link with Dual Queue AQM, the sender access link, in transit, or at a peering link.}
    \label{fig:motivation_topology}
    \vspace{-1em}
\end{figure}

We run a small scale motivating experiment with the topology in Figure~\ref{fig:motivation_topology}  to show why L4S deployment cannot be viewed only as a receiver access problem. We consider four scenarios: an access baseline with no upstream bottleneck and three scenarios that each add a single upstream FIFO bottleneck. The added bottleneck is a 1~Gbps sender access link, a 4~Gbps peering link, or a 4~Gbps core link saturated with bulk background traffic. All scenarios include the 100~Mbps receiver access Dual Queue AQM. These experiments thus represent a setting where the receiver access network supports L4S but another part of the end-to-end path introduces congestion without L4S aware isolation. The L4S flows have a 30~ms base RTT.

\begin{figure*}[t]
    \centering
    \includegraphics[width=\textwidth]{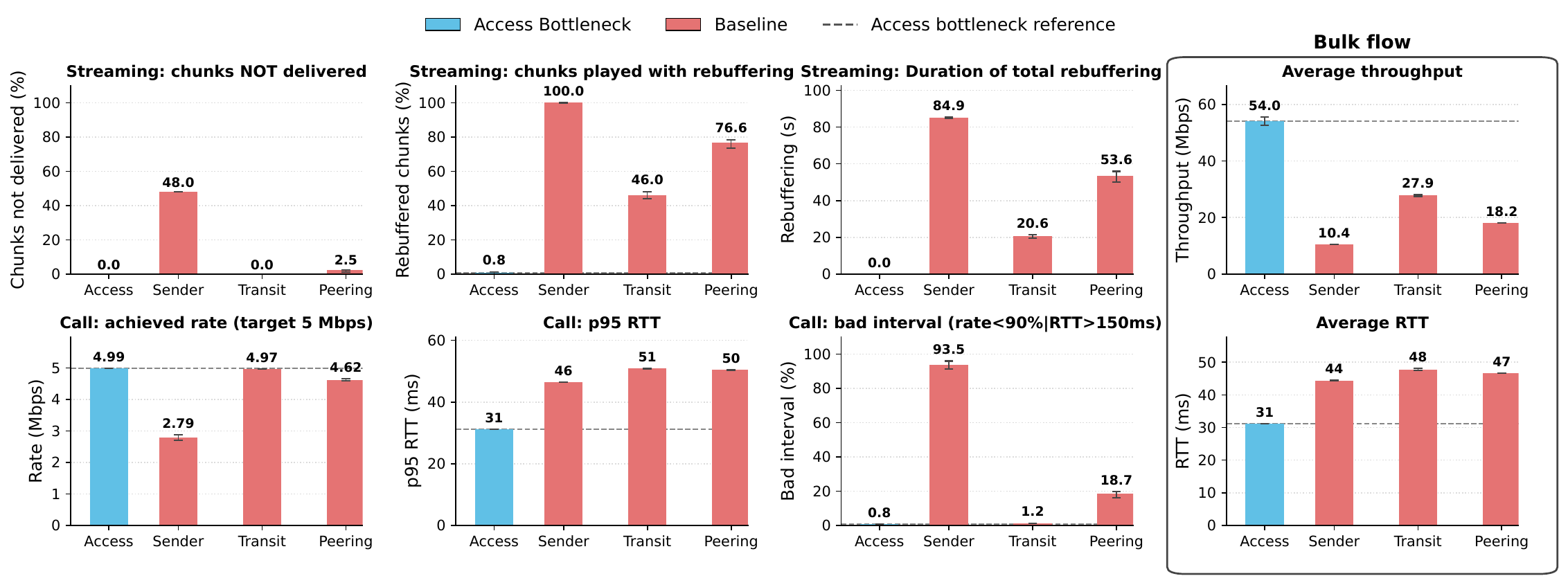}
    \caption{Motivating experiment showing the effect of upstream congestion on L4S performance. The receiver access Dual Queue AQM is present in all scenarios. The first scenario (in blue) is the case when only the receiver access is a bottleneck, while the other scenarios (in red) add an upstream FIFO bottleneck at a peering, sender access, or transit link. We refer to these as baseline values, because they will serve as the reference point for evaluating our proposed solution. The dashed line in the bulk throughput plot shows the approximate equal share rate for the access baseline. Even with L4S support at the receiver access link, upstream FIFO congestion can reduce L4S throughput, increase queuing delay, and degrade application performance.}
    \label{fig:motivation}
\end{figure*}

We evaluate both transport-level and application-level effects. For the bulk-flow experiment, one L4S flow competes with one bulk CUBIC flow (Fig.~\ref{fig:motivation}). The dashed line in the throughput plot shows the approximate 50~Mbps equal-share rate for the access-only baseline; in the other scenarios it serves as an access-only reference rate, not the fair-share capacity of the upstream FIFO links. For the application experiments, five L4S flows emulate video-streaming and video-call traffic while sharing the path with five bulk CUBIC flows. The video-streaming workload uses an encoding rate of 10~Mbps per L4S flow; the video-call workload uses a target rate of 5~Mbps per L4S flow. For video calls, a bad interval is one in which the achieved rate is below 90\% of the target or the RTT exceeds 150~ms. 

Figure~\ref{fig:motivation} shows that the access baseline with Dual Queue AQM delivers as expected: the L4S flow achieves approximately the equal share rate at the receiver access link while maintaining about 1~ms of queuing delay. When an upstream FIFO bottleneck is introduced, however, the L4S flow is no longer protected from classic cross traffic. Average throughput drops from 54~Mbps in the access baseline to 18.2~Mbps, 10.4~Mbps, and 27.9~Mbps in the peering, sender access, and transit congestion scenarios, respectively, while average queuing delay increases by more than 13~ms in all three cases.  

The application results show that this transport degradation directly affects user experience. Upstream congestion introduces substantial rebuffering for video streaming and affects a large fraction of played chunks. For video calls, the sender access scenario is especially harmful, reducing the achieved rate and causing bad intervals for most of the experiment. The peering and transit scenarios also increase queuing delay, which can cause bad video call intervals even when the achieved rate remains close to the target.

The key observation is that L4S performs well only when the congestion point matches the deployment assumption: a receiver access bottleneck with Dual Queue AQM. L4S support at the access link does not protect a flow from congestion earlier in the path. This motivates the main direction of this paper: a practical deployment strategy that extends low latency isolation to multiple congestion points using mechanisms feasible for wide area and interdomain deployment.

\section{Enabling L4S deployment at scale} \label{sec:solution-description}

We propose a deployment strategy for L4S that preserves end-to-end low latency service without requiring Dual Queue AQM beyond the receiver access link. It relies on mechanisms that are familiar to network operators: BGP communities for route signaling, SRv6 for path steering inside participating domains, and DSCP-based priority queuing for local isolation at shared bottlenecks. Rather than requiring universal router upgrades, our design targets a small set of locations where participating networks can apply isolation and routing policies.

Our design makes two major assumptions. First, routers act on endpoint markings rather than inferring application intent; L4S traffic is identified by the ECN field. Second, DSCP markings are not assumed to be preserved end-to-end. Upgraded routers use ECN classification to set a local DSCP value for queue selection within their managed domain. This allows L4S packets to use existing DSCP-based priority queuing mechanisms, rather than requiring every router to implement classification specific to L4S. Packets with this local DSCP are assigned to a low latency queue that has strict priority up to an operator configured rate cap, which prevents starvation of best-effort traffic (see Section~\ref{sec:classic-impact}).

We organize routers into Access, Core, and Peering roles. Access routers sit at the edges of participating ASes, serving sender and receiver roles. The sender access router classifies ECT(1) or Congestion Experienced (CE) packets as L4S, following the L4S rule, and treats all others as classic. It then marks L4S packets with a local DSCP value (Fig.~\ref{fig:dscp_remarking}) for the benefit of downstream routers in the managed domain, and places L4S packets into a low latency queue with a rate cap.

\begin{figure}[h]
\footnotesize
\caption{Remarking of L4S traffic at the sender access router.}
\label{fig:dscp_remarking}

\hrule\vspace{0.2em}

\begin{tabular}{@{}ll@{}}
\multicolumn{2}{@{}l}{\textbf{Before sender access DSCP processing}} \\
Header: & \texttt{IP6 class \textbf{<DSCP: CS0, ECN: ECT(1)>}}\\[-0.15em]

\multicolumn{2}{@{}l@{}}{\rule{\linewidth}{0.4pt}} \\[0.15em]

\multicolumn{2}{@{}l}{\textbf{After sender access DSCP processing}} \\
Header: & \texttt{IP6 class \textbf{<DSCP: AF11, ECN: ECT(1)>}} \\[0.15em]
\multicolumn{2}{@{}l}{\textit{(NH, Flow, Flags, Length unchanged)}}
\end{tabular}

\vspace{0.2em}\hrule

\vspace{0.15em}
\normalfont
\noindent
DSCP field is remarked from CS0 (no marking) to AF11 (local marking for queue selection within the managed domain); ECT(1) bit is preserved.
\par
\vspace{-1em}

\end{figure}

The sender access router further maintains two routing tables, for classic traffic and for low latency paths. Route placement is determined by a BGP community. The receiver access router attaches an L4S community to receiver prefixes reachable through paths that support L4S. ASes that support L4S preserve this community, while other ASes strip it. A route that still carries the community at the sender AS therefore indicates a path advertised as supporting L4S and is installed in the L4S table; routes without it go into the main table (Fig.~\ref{tab:bgp_table_selection}). This mechanism is used as an experimental proof of concept for L4S path signaling between ASes.

\begin{figure}[h]
\footnotesize
\caption{Route installation based on BGP communities at the sender access router.}
\label{tab:bgp_table_selection}

\hrule\vspace{0.2em}

\begin{tabular}{@{}ll@{}}
\multicolumn{2}{@{}l}{\textbf{L4S Route}} \\
\textbf{Community:} & \texttt{\textbf{<L4S-comm>}} \\
\textbf{AS Path:} & \texttt{<path via transit with L4S support>} \\
\textbf{Table:} & \texttt{\textbf{L4S table}} \\[-0.15em]

\multicolumn{2}{@{}l@{}}{\rule{\linewidth}{0.4pt}} \\[0.15em]

\multicolumn{2}{@{}l}{\textbf{Classic Route}} \\
\textbf{Community:} & \texttt{none} \\
\textbf{AS Path:} & \texttt{<path via transit without L4S support>} \\
\textbf{Table:} & \texttt{main} \\[0.15em]
\multicolumn{2}{@{}l}{\textit{(Both routes share the same receiver prefix)}}
\end{tabular}

\vspace{0.2em}\hrule
\vspace{0.15em}

\normalfont
\noindent
The route whose path to the receiver traverses only ASes that support L4S carries the L4S BGP community and is installed into the L4S routing table. The route whose path passed through an AS without L4S support does not carry the community and is placed in the main routing table.
\par

\vspace{-1em}

\end{figure}

The sender access router also steers traffic within the local AS using SRv6 encapsulation. Routes in the L4S and main tables carry different segment lists: L4S packets are directed toward the dedicated core router with L4S support, while classic packets follow the default core path (Fig.~\ref{fig:srv6_encap}). SRv6 adds encapsulation overhead, reducing the effective payload MTU. In our configuration, this overhead is 80 bytes (a 40-byte outer IPv6 header, an 8-byte SRH base header, and two 16-byte SIDs). We therefore reduce the MTU at the sender from 1500 to 1420 bytes so that packets encapsulated with SRv6 remain within the 1500-byte underlay MTU.

\begin{figure}[h]
\footnotesize
\caption{SRv6 encapsulation at the sender access router.}
\label{fig:srv6_encap}

\hrule\vspace{0.2em}

\begin{tabular}{@{}ll@{}}
\multicolumn{2}{@{}l}{\textbf{L4S packet after Access processing}} \\
Outer: & \texttt{IP6 class \textbf{<DSCP: AF11, ECN: ECT(1)>}} \\
Source: & \texttt{<Access-L4S-SID>} \\
Destination: & \texttt{\textbf{<L4S-Core-SID>}} \\
RT6: & \texttt{[0] <Egress-SID>,\; [1] \textbf{<L4S-Core-SID>}} \\
Inner: & \texttt{IP6 class \textbf{<DSCP: AF11, ECN: ECT(1)>}} \\
Flags: & \texttt{\textbf{[.E]}} \\[-0.15em]

\multicolumn{2}{@{}l@{}}{\rule{\linewidth}{0.4pt}} \\[0.15em]

\multicolumn{2}{@{}l}{\textbf{Classic packet after Access processing}} \\
Outer: & \texttt{IP6 (no DSCP/ECN marking)} \\
Source: & \texttt{<Access-Classic-SID>} \\
Destination: & \texttt{\textbf{<Classic-Core-SID>}} \\
RT6: & \texttt{[0] <Egress-SID>,\; [1] \textbf{<Classic-Core-SID>}} \\
Inner: & \texttt{IP6 (no DSCP/ECN marking)} \\
Flags: & \texttt{[.]} \\[0.15em]
\multicolumn{2}{@{}l}{\textit{(Outer NH, Inner NH, Flow unchanged)}}
\end{tabular}

\vspace{0.2em}\hrule

\vspace{0.15em}
\normalfont
\noindent
The \texttt{Destination} and \texttt{RT6} segment list fields differ between L4S and classic packets: L4S traffic is steered toward the dedicated core router with L4S support (\texttt{<L4S-Core-SID>}), while classic traffic is steered toward the standard core router (\texttt{<Classic-Core-SID>}). SID names based on router roles are used in place of raw IPv6 segment addresses.
\par
\vspace{-1em}

\end{figure}

The receiver access router runs Dual Queue AQM to maintain isolation between L4S and classic traffic. When acting as the SRv6 segment endpoint, it follows the End.DT6 behavior: it copies ECN bits from the outer to the inner header before decapsulating, so that any CE marking set in transit is preserved and delivered to the receiver (Fig.~\ref{fig:receiver_access_ce_copy}).

\begin{figure}[h]
\footnotesize
\caption{Receiver access decapsulation with CE propagation.}
\label{fig:receiver_access_ce_copy}

\hrule\vspace{0.2em}

\begin{tabular}{@{}ll@{}}
\multicolumn{2}{@{}l}{\textbf{Encapsulated packet before receiver access decapsulation}} \\
Outer: & \texttt{IP6 class \textbf{<DSCP: AF11, ECN: CE>}} \\
Outer NH: & \texttt{Routing (43)} \\
Source: & \texttt{<L4S-Core-SID>} \\
Destination: & \texttt{<Receiver-Access-SID>} \\
Inner: & \texttt{IP6 class \textbf{<DSCP: AF11, ECN: ECT(1)>}} \\[-0.15em]

\multicolumn{2}{@{}l@{}}{\rule{\linewidth}{0.4pt}} \\[0.15em]

\multicolumn{2}{@{}l}{\textbf{Decapsulated packet after receiver access processing}} \\
Inner: & \texttt{IP6 class \textbf{<DSCP: AF11, ECN: CE>}} \\[0.15em]
\multicolumn{2}{@{}l}{\textit{(NH, Flow unchanged)}}
\end{tabular}

\vspace{0.2em}\hrule

\vspace{0.15em}
\normalfont
\noindent
The outer header carries ECN: CE (set by a congested queue in transit). Before decapsulation, this CE marking is copied to the inner header, so the receiver sees the congestion signal and it is not lost on decapsulation.
\par
\vspace{-1em}
\end{figure}

The receiver access router also originates the BGP advertisements that carry the L4S community, attaching it to receiver prefixes that are reachable through paths with L4S support.

Core routers are internal to a transit AS, rather than facing an access network or sitting at an interdomain boundary. They only implement the SRv6 ``End'' behavior, processing the active SID and forwarding to the next segment. No classification or queue management specific to L4S is required.

Peering routers sit at AS boundaries. Their primary role is managing the BGP L4S community: an AS that supports L4S preserves the community on advertised prefixes, while an AS without L4S support strips it because it cannot guarantee low latency treatment (Fig.~\ref{fig:peering_community}). This allows both the sender access router and downstream peering routers to determine whether a path is advertised as supporting L4S. Like the sender access router, peering routers install prefixes carrying the community into the L4S routing table and others into the main table.

Peering routers also perform ingress and egress functions analogous to access routers. On ingress, they classify packets as L4S or classic, restore DSCP markings cleared in transit, and encapsulate packets whose next hop is within the AS. On egress, they select a routing table based on ECN classification, copy ECN and local DSCP from the outer to the inner header, and decapsulate before forwarding. Peering links shared by L4S and classic traffic can use a low latency queue with a rate cap, providing local isolation at the bottleneck.

The BGP community in our solution is a capability signal rather than a verifiable service guarantee. We assume that participating ASes preserve the community only when they provide the corresponding treatment and remove it otherwise. Misconfiguration or stale policy can violate this assumption, so a production deployment would require agreed propagation rules and operational validation, for example through active monitoring. Similarly, although DSCP is reconstructed locally rather than trusted end-to-end, endpoint ECN markings remain part of the classification trust boundary; networks that do not trust attached endpoints may require additional policing. If the advertised capability cannot be validated, an operator can conservatively treat the route as best effort rather than rely on the advertised low latency capability.

\begin{figure}[h]
\footnotesize
\caption{How peering routers handle L4S capability signaling across ASes with and without L4S support.}
\label{fig:peering_community}

\hrule\vspace{0.2em}

\begin{tabular}{@{}p{0.48\linewidth}p{0.48\linewidth}@{}}
\multicolumn{2}{@{}l}{\textbf{AS with L4S support}} \\[0.1em]
\textbf{Entering Peering} & \textbf{Leaving Peering} \\[0.1em]
\begin{tabular}[t]{@{}l@{}}
\texttt{NLRI: <L4S-prefix>} \\
\texttt{COMMUNITY: \textbf{<L4S-comm>}}
\end{tabular}
&
\begin{tabular}[t]{@{}l@{}}
\texttt{NLRI: <L4S-prefix>} \\
\texttt{COMMUNITY: \textbf{<L4S-comm>}}
\end{tabular}
\end{tabular}

\vspace{0.05em}\rule{\linewidth}{0.4pt}\vspace{0.15em}

\begin{tabular}{@{}p{0.48\linewidth}p{0.48\linewidth}@{}}
\multicolumn{2}{@{}l}{\textbf{AS without L4S support}} \\[0.1em]
\textbf{Entering Peering} & \textbf{Leaving Peering} \\[0.1em]
\begin{tabular}[t]{@{}l@{}}
\texttt{NLRI: <L4S-prefix>} \\
\texttt{COMMUNITY: \textbf{<L4S-comm>}}
\end{tabular}
&
\begin{tabular}[t]{@{}l@{}}
\texttt{NLRI: <L4S-prefix>} \\
\texttt{COMMUNITY: \textbf{not present}}
\end{tabular}
\end{tabular}

\vspace{0.2em}\hrule

\vspace{0.15em}
\normalfont
\noindent
An AS with L4S support preserves the L4S BGP community on the prefix reachable through a path that supports L4S (\texttt{<L4S-prefix>}) and passes it on unchanged. An AS without L4S support strips the community. AS\_PATH, and the classic prefix and its communities, are identical in all four cases and are omitted. The sender access router uses the presence or absence of this community to determine whether the full path to a prefix supports L4S.
\par
\vspace{-1em}

\end{figure} 

\section{Experiment methodology} \label{sec:methodology}

We evaluate our deployment strategy through a series of systematic experiments that vary the bottleneck location on the network path and the set of enabled mechanisms.

\textbf{Experiment platform:} We conduct all experiments on FABRIC~\cite{baldin2019fabric}, a national programmable networking testbed. Each node in our experiment is a virtual machine running Ubuntu 22.04, with 4~cores and 4~GB~RAM. 

\begin{figure*}[t]
    \centering
    \begin{subfigure}[t]{0.44\textwidth}
        \vspace{0pt}
        \centering
        \includegraphics[width=\linewidth]{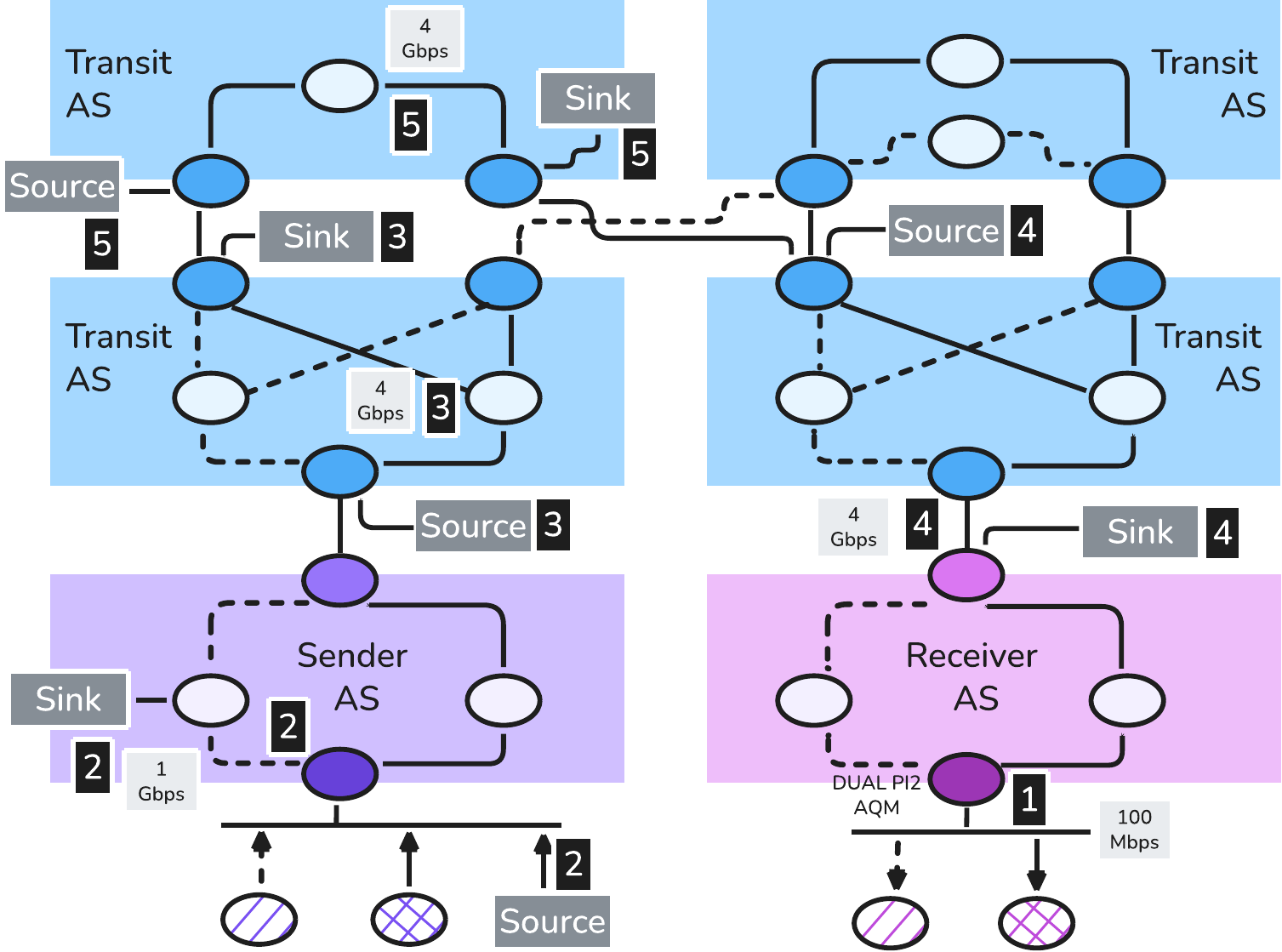}
        \caption{Experiment topology used in our FABRIC evaluation. Solid (dashed) lines denote the intended path of classic (L4S) traffic.}
        \label{fig:large_topology}
    \end{subfigure}\hfill
    \begin{subfigure}[t]{0.49\textwidth}
        \vspace{0pt}
        \centering
        \footnotesize
        \setlength{\extrarowheight}{1.8pt}
        \begin{tabular}{C{0.17\textwidth}|L{0.25\textwidth}|L{0.35\textwidth}}
        \hline
        \textbf{\shortstack[c]{Heavy traffic\\path}} & \textbf{\shortstack[c]{Where is the\\bottleneck?}} & \textbf{Proposed mechanism at bottleneck} \\
        \hline
        \sqnum{1} & Receiver access link & Existing DualPI2 at receiver access (no additional mechanism) \\
        \hline
        \sqnum{2} & Sender access link & DSCP-based L4S priority queue with rate cap \\
        \hline
        \sqnum{3} & Transit AS with L4S support & SRv6 steering to isolated internal path \\
        \hline
        \sqnum{4} & Receiver peering link & DSCP-based L4S priority queue with rate cap \\
        \hline
        \sqnum{5} & Legacy transit AS & BGP L4S community + SRv6 steering to avoid legacy path \\
        \hline
        \sqall & Multiple bottlenecks (all locations) & Full combination of mechanisms from Scenarios 2--5 \\
        \hline
        \end{tabular}
        \captionof{table}{Experiment scenarios and the mechanism used in our proposed solution for each bottleneck location.}
        \label{tab:scenarios}
    \end{subfigure}
    \caption{Evaluation topology and bottleneck scenarios.}
    \label{fig:topology_scenarios}
    \vspace{-1em}
\end{figure*}

\textbf{Topology:} Figure~\ref{fig:large_topology} shows our network topology. The topology has six autonomous systems: a sender domain, a receiver domain, and four transit domains. The first transit domain on the sender side and the last transit domain before the receiver are connected by two parallel paths. One path represents a transit network with L4S support, while the other represents a legacy transit network without L4S support. Each AS has at least one edge router for interdomain peering. Each AS with L4S support also has two parallel core routers, Core1 and Core2. This allows traffic inside the AS to be steered either through an internal path isolated for L4S or through the default internal path. The sender and receiver domains also include access routers connected to the end hosts. The L4S sender and receiver connect to the access routers in their respective domains. A separate classic sender and receiver pair also connects to the same access routers and uses the same source and destination domains. In addition, we attach background traffic pairs around specific potential bottlenecks. The path of each background pair is fixed using static IPv6 routes before each trial, so that the background traffic always crosses the intended upstream bottleneck.

The topology does not reproduce any specific production ISP. Rather, it abstracts the path elements relevant to incremental L4S deployment: sender and receiver access networks, interdomain peering, L4S and legacy transit domains, and alternative intradomain paths. This allows us to independently place congestion at each major bottleneck class while evaluating both intradomain steering and interdomain signaling.

\textbf{Network settings:} We configure five possible bottleneck locations along the end-to-end path. These queues are present in every scenario; the scenarios differ in which location is loaded with background traffic.

\begin{itemize}[noitemsep, topsep=0pt]

  \item \textbf{Receiver access:} 100~Mbps DualPI2 bottleneck on the foreground traffic path using \texttt{tc-dualpi2}~\cite{l4srepo}.

  \item \textbf{Sender access:} 1~Gbps queue. The baseline uses a single \texttt{tc-bfifo} FIFO; our design uses strict priority with DSCP-classified L4S and best-effort queues, capping L4S at 200~Mbps to protect best-effort traffic. Both queues run PIE with ECN, targeting 1~ms for L4S and 15~ms for best effort. Because L4S load remains below the cap, its queue records no marks or drops; Section~\ref{sec:microbench} separately evaluates the design under saturation.

  \item \textbf{Transit core with L4S support:} 4~Gbps single FIFO queue at a core router in the transit AS with L4S support.

  \item \textbf{Legacy transit core:} 4~Gbps single FIFO queue at a core router in the legacy transit AS.

  \item \textbf{Receiver peering link:} 4~Gbps queue at the peering link. By default, this link uses a single FIFO queue. As part of our proposed deployment strategy, it can also use the same priority scheduler with two queues as the sender access link, but with the L4S queue capped at 1~Gbps.
\end{itemize}

The base RTT on the end-to-end path between the sender and receiver end hosts is 30~ms, applied with \texttt{netem} at the receiver side. This applies to both sender-receiver pairs (L4S and classic). The background traffic pairs use a 1~ms base RTT. We use \texttt{tc-htb} for shaping, \texttt{tc-bfifo} for the default FIFO queues, \texttt{tc-dualpi2} at the receiver access, and \texttt{tc-prio} + \texttt{tc-htb} + \texttt{tc-pie} for the two-queue priority configuration. Queue limits at each bottleneck are sized at 2~BDP.

\begin{figure*}[t]
    \centering
    \begin{subfigure}[t]{0.33\textwidth}
        \centering
        \includegraphics[width=\linewidth]{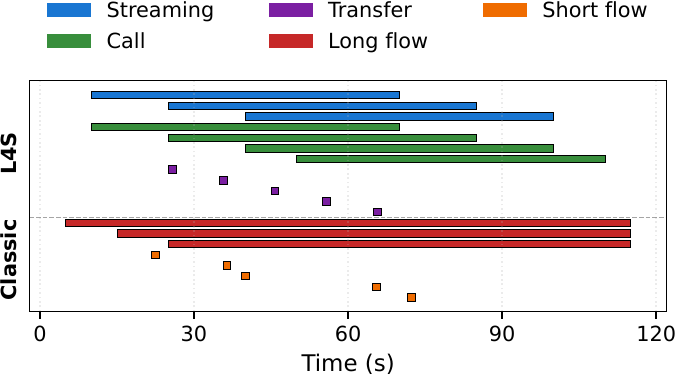}
        \caption{Foreground (L4S + classic).}
        \label{fig:fg}
    \end{subfigure}\hfill
    \begin{subfigure}[t]{0.33\textwidth}
        \centering
        \includegraphics[width=\linewidth]{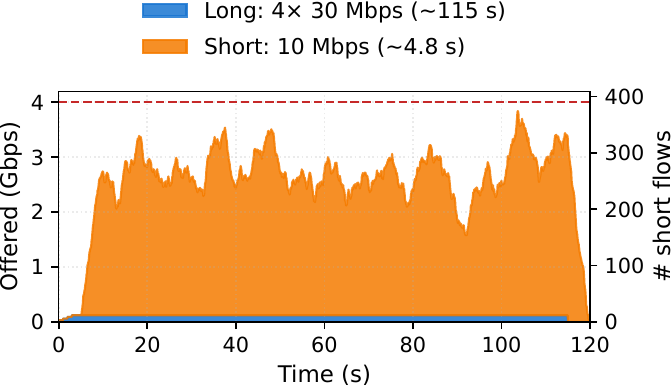}
        \caption{Heavy background: 4~Gbps core/peering.}
        \label{fig:bg-heavy-4g}
    \end{subfigure}\hfill
    \begin{subfigure}[t]{0.33\textwidth}
        \centering
        \includegraphics[width=\linewidth]{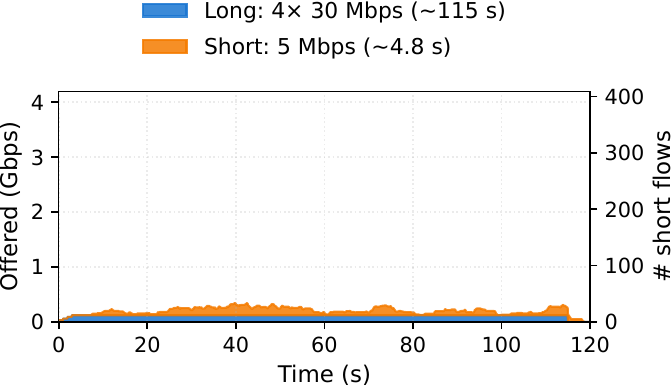}
        \caption{Light background: non-bottleneck pairs.}
        \label{fig:bg-light}
    \end{subfigure}

    \caption{\textbf{Workload schedule for a 120\,s experiment.}
    (a) Foreground traffic includes three 10\,Mbps TCP Prague video streams, four 5\,Mbps UDP Prague video calls, five 10\,MiB TCP Prague file transfers, three long-lived TCP CUBIC bulk flows, and five 5\,MiB TCP CUBIC file transfers.
    (b)--(c) Background traffic demand. The heavy 4\,Gbps core/peering profile uses four 30\,Mbps long flows and 10\,Mbps short flows lasting about 4.8\,s. The light non-bottleneck profile uses four 30\,Mbps long flows and 5\,Mbps short flows lasting about 4.8\,s. The left axis shows offered load, and the right axis shows active short flows.}

    \label{fig:workload}
\end{figure*}

\textbf{Routing:} All routers run FRRouting (FRR) with BGP and OSPFv3, with IPv6 forwarding enabled. We implement the mechanisms from Section~\ref{sec:solution-description} using FRR route-maps and Linux SRv6; each mechanism can be independently enabled or disabled to isolate its effect.

\textit{BGP L4S community.} We use community \texttt{65535:100} to mark paths that support L4S. The receiver access router attaches it to receiver prefixes. In ASes with L4S support, edge routers match this community on inbound routes, install matching prefixes in table~100, and assign \texttt{local-preference 200}, making these paths preferred for L4S traffic. The legacy transit domain strips the community on inbound routes and is configured as the default path for classic traffic by assigning \texttt{local-preference 150} to routes whose AS path matches that domain. When this mechanism is disabled, no community is attached and all traffic takes the default path through the legacy domain.

\textit{SRv6 steering within the AS.} OSPFv3 link costs make Core2 the default internal path (cost 10 toward Core2, cost 20 toward Core1). To steer L4S traffic onto the isolated Core1 path, edge and access routers use an \texttt{ip6tables} mangle rule to mark packets with ECT(1) or CE, then an IPv6 policy rule directs those packets to table~100, which holds SRv6 \texttt{encap} routes via Core1 and the next BGP hop. Core routers apply the SRv6 End behavior on their loopbacks; edge and access routers install an End.DT6 entry to decapsulate at the end of the segment list. When SRv6 steering is disabled, the encapsulation routes, marking rule, and policy rule are removed, and L4S traffic falls back to the Core2 default path.

\textbf{Workload:} Fig.~\ref{fig:workload} visualizes the workload in the network. The end-to-end workload includes both L4S and classic traffic. We use MGEN~\cite{mgen} to emulate L4S application and background traffic patterns, not full application logic. Classic bulk transfers are generated using \texttt{iperf3} with TCP CUBIC.

The L4S sender runs three traffic classes simultaneously over IPv6: three emulated TCP Prague video streams divided into chunks, four emulated UDP Prague video calls, and five short TCP Prague bulk transfers. Each video stream has a 10~Mbps encoding rate, 60~s duration, 2~s chunks, an initial buffer of two chunks, and a 5~s startup delay. Each video call runs at 5~Mbps, and each short L4S bulk transfer is 10~MiB; the transfers start at staggered times (Fig.~\ref{fig:fg}).

The classic sender runs three persistent CUBIC bulk flows together with five short CUBIC transfers of 5~MiB each, with randomized start times.

In addition to this application traffic, four background pairs of senders and receivers run continuously in every scenario. The pair that crosses the intended upstream bottleneck uses a \emph{heavy} profile, while the other three pairs use a \emph{light} profile. The light pairs keep all links active without saturating them. Each background profile consists of persistent TCP flows with rate limits and short flows with finite sizes. The persistent flows represent steady traffic from content providers, while the short flows represent HTTP object fetches. Among the short flows, 75\% use periodic message pacing, while 25\% use Poisson message arrivals with exponential gaps at the same average rate. This mixes paced and bursty traffic at the aggregate. All flows have application layer rate limits.

The \emph{heavy\_4g} profile (Fig.~\ref{fig:bg-heavy-4g}) loads the 4~Gbps upstream bottlenecks with four persistent 30~Mbps flows and compound-Poisson short-flow bursts.  Events arrive at 10 per second; each generates $\mathrm{Poisson}(5.5)$ flows, capped at 25 and jittered by $\pm 0.25$~s. Each flow sends at 10~Mbps for approximately 4.8~s. The 120~Mbps base load plus bursts occasionally exceeds link capacity, producing natural saturation events. Each trial generates approximately 6{,}600 short flows on the heavy background pair.

The \emph{heavy\_1g} profile (not shown) is used for the 1~Gbps sender access bottleneck. It has the same structure as \emph{heavy\_4g} with lower rates per flow: four persistent flows at 15~Mbps each, and short flow events at rate 3.5 per second with batch size drawn from $\mathrm{Poisson}(5.0)$, capped at 20. Each short flow is 8~Mbps and lasts around 4.8~s. This produces an average offered load of $\sim$700~Mbps, with occasional random bursts above the 1~Gbps link capacity. A single trial generates approximately 2{,}100 short flows on the heavy background pair.

The \emph{light} profile (Fig.~\ref{fig:bg-light}) keeps all links active, but not saturated: four persistent flows at 30~Mbps each, and 1.0 short flow events per second with batch size $\mathrm{Poisson}(4.0)$, capped at 15. Each short flow is 5~Mbps and lasts approximately 4.8~s. This yields
$\sim$220~Mbps average offered load per pair with low utilization on the pairs away from bottlenecks.

All background senders and receivers are killed and restarted for each trial, so no state carries over.

\textbf{Scenarios:} Fig.~\ref{tab:scenarios} summarizes the scenarios we evaluate. Each scenario is identified by the numbered pair of source and sink nodes (Figure~\ref{fig:large_topology}) that carries the heavy background traffic. In a given bottleneck scenario, we send the heavy background traffic between the pair bearing that scenario number, which is placed so that the traffic crosses the intended upstream bottleneck. Scenario~\sqnum{1} is the access baseline: only the receiver access DualPI2 bottleneck is active and no upstream bottleneck is stressed. The next scenarios each add one upstream bottleneck and compare the case without our proposed mechanism to the case where the relevant mechanism is enabled: \sqnum{2} sender access, \sqnum{3} transit core with L4S support, \sqnum{4} receiver peering link, and \sqnum{5} legacy transit core. Finally, scenario~\sqall evaluates the stress case where all bottleneck locations carry heavy traffic simultaneously, both without and with the full set of proposed mechanisms. The relevant mechanism depends on the bottleneck location: at peering and sender access links we use the DSCP-based priority queue; inside the transit AS with L4S support we use SRv6 to steer L4S traffic onto the isolated internal path; and for the legacy transit AS we use the BGP L4S community together with SRv6 steering so that L4S traffic avoids the legacy transit path.

\textbf{Metrics:}
At the \emph{transport level}, we record per-flow throughput and queuing delay. For TCP Prague video-streaming and bulk-transfer flows, these metrics are collected from socket statistics. For UDP-Prague video-call flows, they are collected from the Prague feedback information stream. At the \emph{router level}, we sample queue length, ECN marks, and drop counts at each instrumented bottleneck throughout the trial. At the \emph{application level}, we report L4S streaming, video-call, and short-transfer metrics, as well as classic throughput, RTT, and flow completion time (FCT). A \emph{bad interval} for a video call is a 100~ms interval in which the mean achieved rate is below 90\% of the target rate or the mean RTT exceeds 150~ms.

\textbf{Trials:} Each scenario is repeated for eight independent 120~s trials, across which we report mean and Q25/Q75 interquartile range.

\begin{figure*}[t]
    \centering
    \includegraphics[width=\textwidth]{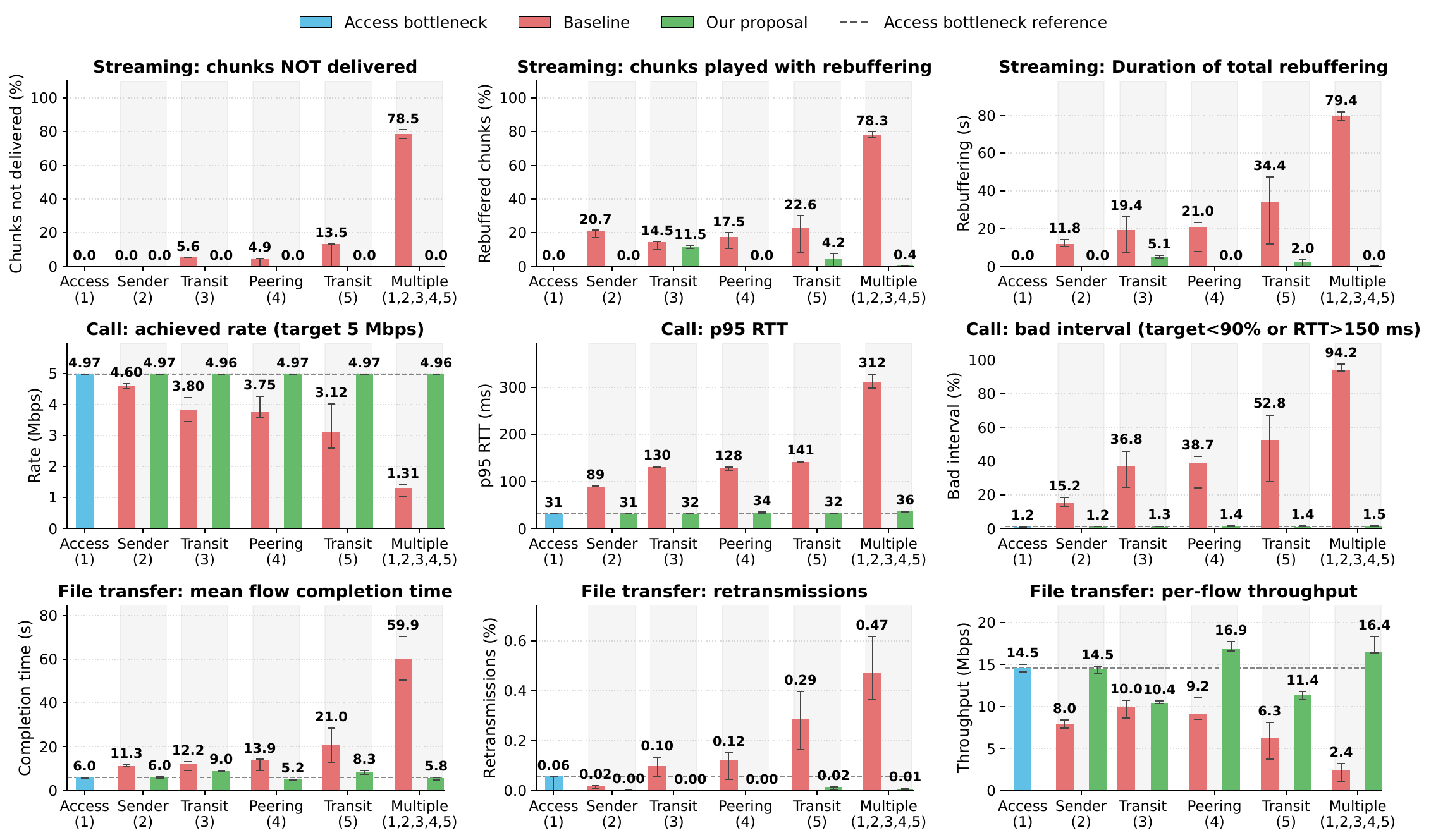}
    \caption{L4S application performance across the eleven scenarios.
    The rows correspond to video streaming, video calling, and short bulk transfers.
    Error bars show the 25th--75th percentile range across eight trials.
    Blue shows the access baseline. Red shows scenarios in which an upstream bottleneck is active but the corresponding solution mechanism is not enabled. Green denotes scenarios in which the corresponding solution mechanism is enabled.}
    \label{fig:all-scenarios}
\end{figure*}

\section{Experiment results}
\label{sec:results}

Figure~\ref{fig:all-scenarios} summarizes L4S application performance across the eleven scenarios in Fig.~\ref{tab:scenarios}. All scenarios include the receiver access DualPI2 bottleneck: the blue bar is the access baseline, red bars add an upstream bottleneck without the corresponding mechanism, and green bars show performance with the mechanism enabled.

\subsection{Recovery by mechanism}
\label{sec:mechanism-recovery}

\textbf{Each proposed mechanism recovers L4S performance when the
corresponding upstream bottleneck is isolated.} We use the
single bottleneck scenarios in Fig.~\ref{fig:all-scenarios} to isolate the effect of each mechanism.

\paragraph{Priority queuing protects shared access and peering bottlenecks} Priority queuing applies when L4S and classic traffic share a domain border link. We evaluate it in scenarios \sqnum{2} and \sqnum{4}: in \sqnum{2}, heavy background traffic induces congestion at the sender access router; in \sqnum{4}, it induces congestion at the receiver peering link. Without priority queuing, the sender access bottleneck causes 12~s of streaming rebuffering, 89~ms call p95 RTT, and 11~s mean transfer time; the peering bottleneck causes 21~s of rebuffering, 128~ms call p95 RTT, and 14~s mean transfer time. Enabling priority queuing brings both close to the access baseline: at sender access, rebuffering drops to zero, call p95 RTT to 31~ms, and mean transfer time to 6~s; at peering, rebuffering drops to zero, call p95 RTT to 34~ms, and short transfers finish in about 5~s.

\paragraph{SRv6 steering isolates L4S traffic inside an upgraded transit AS} In scenario \sqnum{3}, heavy traffic congests the transit core. Without SRv6, L4S traffic shares that path, causing about 19~s of streaming rebuffering and 130~ms call p95 RTT. Enabling SRv6 steers L4S onto the isolated path, restoring call p95 RTT to 32~ms and mean transfer time to 9~s. Streaming does not fully return to the access baseline: about 5~s of rebuffering remains, which we attribute to bursty classic traffic at the downstream DualPI2 bottleneck coupling into the L4S queue. SRv6 helps at the upstream core bottleneck but cannot eliminate downstream interactions at the receiver access link.

\paragraph{BGP community signaling avoids legacy transit} In scenario \sqnum{5}, heavy traffic congests the legacy transit core. Without BGP community signaling, L4S traffic is routed through this core, causing about 34~s of streaming rebuffering, 141~ms call p95 RTT, 3~Mbps call achieved rate, and 21~s mean transfer time. Enabling the L4S BGP community reroutes L4S traffic onto the transit path with L4S support, bringing performance close to the access baseline: streaming rebuffering drops to 2~s, call p95 RTT to 32~ms, call achieved rate to 4.97~Mbps, and mean transfer time to 8~s.

\subsection{Full solution deployment under simultaneous bottlenecks}
\label{sec:combined}

\textbf{When all upstream bottlenecks are loaded simultaneously, the full deployment recovers L4S performance. L4S performance collapses without the proposed mechanisms.}

\begin{figure*}[t]
    \centering
    \includegraphics[width=\textwidth]{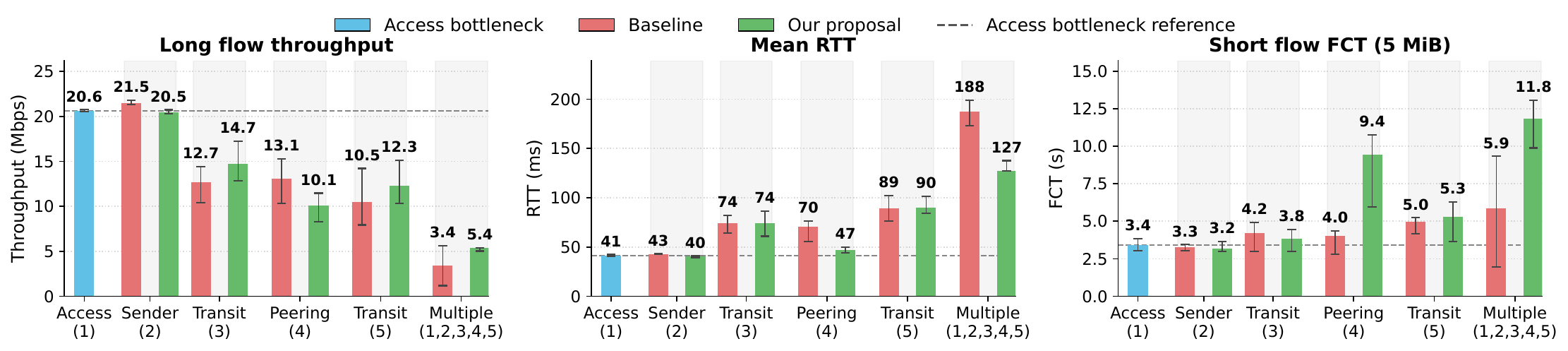}
    \caption{Classic CUBIC performance: long-flow throughput (left), mean RTT (center), and 5~MiB short-flow completion time (right). Blue/dashed lines mark the access-bottleneck baseline; red/green compare each mechanism disabled/enabled. Whiskers show 25th--75th percentiles over eight trials. Sender Bottleneck affects only the L4S path, so classic metrics remain at baseline.}
    \vspace{-0.1cm}
    \label{fig:classic}
\end{figure*}

In the stress scenario, all upstream bottlenecks are loaded simultaneously. Without mitigation, L4S performance collapses: streaming flows incur 79~s of rebuffering, video calls achieve 1.3~Mbps with 312~ms p95 RTT, and short transfers average 60~s, versus 6~s in the access baseline. Full deployment combines priority queuing at sender access and peering links, BGP L4S community signaling, and SRv6 path steering, restoring performance to zero rebuffering, 36~ms call p95 RTT, a 4.96~Mbps call rate, and 5.8~s mean transfer time.

\subsection{Effect on classic foreground traffic}
\label{sec:classic-impact}

\textbf{The proposed mechanisms preserve steady throughput for classic traffic, but strict priority isolation can substantially increase completion times for short classic transfers.}

We evaluate CUBIC traffic to determine whether L4S latency gains come at the expense of classic flows. Fig.~\ref{fig:classic} summarizes long-flow throughput, mean RTT, and short-flow FCT across all scenarios. Long CUBIC flows are not starved: in the all-heavy scenario, the full deployment increases throughput from 3.4~Mbps to 5.4~Mbps and reduces mean RTT from 188~ms to 127~ms. Across single-bottleneck scenarios, enabling each mechanism leaves long-flow throughput unchanged and maintains or reduces RTT. 

The tradeoff arises for short CUBIC transfers under strict-priority queuing: FCT increases from 4.0~s to 9.4~s at the peering bottleneck and from 5.9~s to 11.8~s in the all-heavy scenario. These transfers are more sensitive because they spend much of their lifetime in slow start. Operators could choose a different operating point in this trade-off: strict priority could be replaced by a weighted scheduler such as WRR, which favors L4S without granting absolute priority, or by a scheduler that guarantees the classic queue a minimum service share. Operators could also lower the L4S rate cap or adapt it to demand under high L4S load. These choices could protect short classic flows at the cost of weaker L4S isolation or reduced L4S capacity, and may require additional tuning. We therefore treat strict priority as one point in a broader scheduler design space and leave systematic comparison to future work.

\subsection{Microbenchmark: priority queue at saturation}
\label{sec:microbench}

\begin{figure}[h]
    \centering
    \includegraphics[width=0.8\columnwidth]{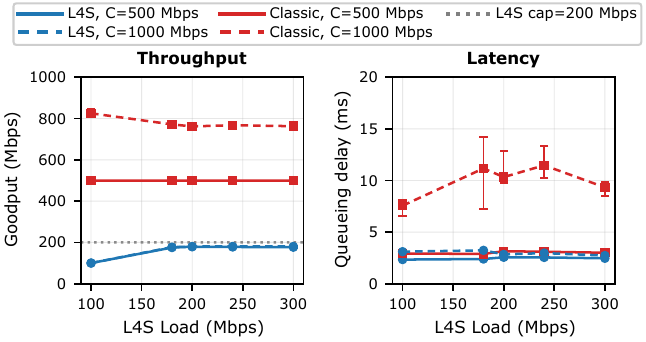}
    \caption{Priority-queue microbenchmark. Color indicates traffic type, and linestyle indicates the classic offered load. Whiskers show Q25/Q75 across trials.}
    \label{fig:microbench}
    \vspace{-0.5cm}
\end{figure}

\textbf{The priority queue keeps L4S delay bounded as L4S approaches its service cap, while classic traffic continues to receive its expected service.}
In the deployment scenarios, the L4S load remains well below the priority-class rate cap, so the benefit of the priority queue comes from strict-priority isolation rather than active ECN marking. To evaluate behavior when the L4S queue itself saturates, we stress-test it on a single-router 1~Gbps bottleneck with a 200~Mbps L4S queue.
For this microbenchmark the queue AQM is CoDel with the \texttt{ce\_threshold} option. Preliminary tests showed that PIE's periodic delay-based probability update did not provide the shallow, deterministic marking needed here; CoDel's \texttt{ce\_threshold} marks every packet whose sojourn time exceeds a fixed threshold, which is closer to the intended L4S queue behavior. This choice does not affect the deployment-scenario results, where the L4S queue never builds a persistent queue.
Both queues use CoDel with ECN marking: 1~ms \texttt{ce\_threshold} on the L4S queue and 15~ms target on the classic queue. We sweep L4S offered load from 100~Mbps to 300~Mbps under two classic loads---500~Mbps (below residual capacity) and 1000~Mbps (overloads classic queue)---and run five 60~s trials at each point.

As shown in Fig.~\ref{fig:microbench}, L4S goodput tracks offered load up to about 175~Mbps and then levels off below the 200~Mbps cap: as the queue approaches saturation, ECN marking causes Prague to reduce its rate rather than letting the queue grow. L4S average queueing delay stays bounded at roughly 2.5--3~ms across the sweep under both classic loads.
Classic traffic is not starved: at 500~Mbps classic load, classic traffic receives its full offered rate. At 1000~Mbps, it converges to the residual capacity left by the capped L4S queue, around 800~Mbps once the L4S queue saturates, with average queueing delay around 10~ms, below the 15~ms classic-queue threshold.

\section{Deployment considerations and limitations}
\label{sec:limitations}

\textbf{Deployment alternatives and operational cost.} The mechanisms evaluated in this work represent one practical
realization of the functions needed to extend L4S service, rather than
the only possible design. For example, other mechanisms could provide interdomain capability signaling or intradomain traffic engineering, and an operator could deploy queues compatible with L4S on shared core paths instead of steering L4S traffic onto an isolated path. Our design favors mechanisms already
available in operator networks and reduces the need for queue changes specific to L4S throughout the core. This choice nevertheless introduces its own operational costs. In particular, path-based isolation requires sufficient path diversity and
capacity, together with configuration and maintenance of the
corresponding steering policies. An operator therefore faces a tradeoff
between provisioning an alternative path for L4S traffic and upgrading
routers on a shared path to provide treatment compatible with L4S. Failures
can further reduce the available isolated capacity or require traffic to
fall back to a path without low latency treatment. We do not evaluate
capacity planning or failure recovery in this work.

\textbf{Interdomain signaling}. Our proposed solution uses BGP communities as one practical mechanism for signaling L4S capability between ASes. This choice is not fundamental to the architecture, but provides an experimental realization of the required interdomain signaling function. Other mechanisms could be used instead; for example, BGP Colored Prefix Routing (CPR) uses Color Extended Communities to associate prefixes with routing intent~\cite{rfc9723-bgp-colored-prefix-routing}.

\textbf{Queue configuration.} The L4S rate cap and priority scheduling parameters are policy choices configured by the operator. A cap that is too low can unnecessarily constrain L4S demand, while a cap that is too high reduces the capacity protected for classic traffic. The values used in our experiments are specific to the evaluated links and workloads; production deployments would need to select and potentially adapt these values according to traffic demand and service objectives.

\textbf{Scalability.} Our design concentrates classification and policy enforcement at access and peering routers, while core routers perform standard SRv6 forwarding. Similarly, the interdomain capability signal is associated with advertised routes rather than individual transport flows. These properties avoid additional state for each flow in the core. However, we do not evaluate control plane scaling with Internet-size routing tables, large numbers of participating ASes, or frequent updates.

\textbf{Evaluation scope.} Our FABRIC experiments provide a controlled and reproducible evaluation in which bottleneck location and deployment mechanisms can be varied independently. They demonstrate the feasibility of the proposed mechanisms and quantify their effects under the evaluated scenarios spanning multiple domains, but do not capture the full diversity of production Internet environments. We do not evaluate heterogeneous RTTs, BBR or unresponsive competing traffic, different L4S adoption ratios, sustained overload across the deployment scenarios, microbursts at production scale, asymmetric routing, frequent routing changes, or link and router failures. These factors may affect the quantitative results and remain important directions for future evaluation.

\section{Conclusion}
\label{sec:conclusion}

L4S is commonly deployed and evaluated with the receiver access link as the main bottleneck. However, our results show that this assumption is not sufficient for end-to-end low latency service. Even when the receiver access link uses Dual Queue AQM, upstream congestion at sender access links, peering links, or transit cores can degrade L4S throughput, increase delay, and affect application performance.

To address this problem, we proposed a practical deployment strategy that extends L4S isolation beyond the access link without requiring every router on the path to implement Dual Queue AQM. The design combines low latency queues with rate caps at shared bottlenecks, SRv6 steering inside upgraded domains, and BGP community signaling across domains. Our experiments show that these mechanisms can recover L4S performance at their targeted bottlenecks and that the full deployment strategy preserves low latency application performance when multiple upstream bottlenecks are active. As with any incremental deployment, the approach requires operator participation, coordination of interdomain signaling, and careful configuration of the low latency queue and its rate cap. Overall, these results suggest that L4S service can be extended end-to-end by combining multiple deployable mechanisms in today's networks.

\section*{Acknowledgment}
This work was supported by the New York State Center for Advanced Technology in Telecommunications (CATT) and by the NSF under awards 2148309 and 2552015.

\clearpage
\bibliographystyle{IEEEtran}
\bibliography{refs}

\end{document}